\documentstyle{article}

\title{\bf The bootstrap for impact factors and the gluon wave function}
\author{Mikhail Braun$^{a)}$, and Gian Paolo Vacca$^{b)}$  \\
$^{a)}$ Department of high-energy physics, University of S. Petersburg\\
$^{b)}$II. Institute f\"ur Theoretische Physik, Universit\"at Hamburg.}
\def\beq{\begin{equation}}
\def\eeq{\end{equation}}
\def\noi{\noindent}
\def\oq{\omega(q)}

\def\eq{\eta (q)}
\def\ea{\eta (q_{1})}
\def\eb{\eta (q_{2})}

\def\xa{\xi(q_1)}
\def\xb{\xi(q_2)}

\def\Ga{{\rm\Gamma}}
\def\ps{{\rm\psi}}
\def\qpa{{q'_1}^2}
\def\qpb{{q'_2}^2}
\begin{document}
\maketitle
\medskip
\noi{\bf Abstract.}
Using the results recently obtained for the non-forward quark and gluon
impact factors, it is shown that their form in the gluon colour channel is
consistent with the ``third bootstrap condition'', namely, that they 
should be proportional to the gluon wave function.
The gluon wave function found from this assumption is 
used to write the full bootstrap condition for the
gluonic potential in the next-to-leading order.

\section{ Introduction.}

Recently the 2nd order corrections were calculated for the 
gluon and quark non-forward impact factors [1,2]. The authors checked 
the so-called ``second bootstrap condition'' for them and found
that it was satisfied. There is however still another (``third'')
bootstrap condition for the impact factors discussed in [3]. 
It is not operative for the elastic particle-particle (PP) amplitudes 
but becomes essential if one considers particle-reggeon (PR) amplitudes.
Such amplitudes naturally appear if one considers inelastic
amplitudes with production of one or several gluons.
Fulfillment of the third bootstrap condition seems necessary
to satisfy unitarity for inelastic amplitudes [4]

As discussed in [3], the third bootstrap condition is just a
requirement that in the gluonic channel
the impact factor for ANY particle,
considered as a function of reggeonic momenta, coincide
with the gluon wave function up to a factor which may depend on the
transferred momentum. In this note we check this requirement for
the impact factors calculated in [1,2].

From the start it is clear that since the gluon wave function is
unknown in the next-to-leading order (NLO), checking of the third
bootstrap condition strictly speaking reduces to verifying that both 
the gluon and quark impact factors have the same dependence on the 
reggeonic momenta. Our result is that this prediction of the 
``third bootstrap'' is true: the dependence on the reggeonic momenta 
is indeed identical for the gluon and quark impact factors.
Assuming that the ``third bootstrap'' condition is valid,
we are able to find the 2nd order corrections to the gluon wave
function and consequently write the main bootstrap condition
in the 2nd order. This is a stringent test for the 2nd order
potential, still unknown at present.

In [5] an ansatz was proposed for the 
gluon trajectory, wave function and  interaction potential to
introduce the running coupling to all orders of the fixed one.
It solves the main bootstrap condition 
(the ``first'' one)  identically.
 It was found in [3] that a part of the
NLO interaction in the gluonic channel which comes from an intermediate
$q\bar q$ pair found from this ansatz coincides with the one found
by direct calculation in [6]. So  the ansatz works
for the part due to an intermediate $q\bar q$ pair.
However our results show that, if the third bootstrap is fulfilled,
the ansatz is not valid for the part of the gluon wave function
generated by  gluonic intermediate states, although it correctly
reproduces the running of the coupling.

\section{The third bootstrap condition}
Consider the non-forward elastic PP amplitude generated by the
exchange of two reggeized gluons in the gluon colour channel.
Its absorptive part
as a function of the angular momentum $j$ and
in the leading and next-to-leading orders can be written as
\beq
{\cal A}_j=\langle\Phi_p|G(E)|\Phi_t\rangle.
\eeq
Here the  functions $\Phi_{p,t}(q,q_1)$ are the impact factors
which represent
the coupling of the external particles ($t$ from the target and $p$
from the projectile) to the two exchanged reggeized 
gluons with momenta $q_1$ and
$q_2=q-q_1$, where $q$ is an overall momentum transfer. 
The 2-gluon Green function $G(E)=(H-E)^{-1}$ with $H$ the 
Hamiltonian and $E=1-j$, can be represented via
the orthonormalized solutions of the homogeneous Schr\"odinger 
equation
\beq
H\Psi_n=E_n\Psi_n.
\eeq
As a result, 
the absorptive part ${\cal A}_j$ can be written as
\beq
{\cal A}_j=
\sum_n\frac{\langle\Phi_p|\Psi_n\rangle\langle\Psi_n|\Phi_t
\rangle}{E_n-E}.
\eeq
We shall be also interested in the PR amplitudes, 
defined as
\beq
\Psi(E)=G(E)|\Phi\rangle,
\eeq
where $\Phi$ is the particle impact factor. Amplitude $\Psi$
satisfies an inhomogeneous Schroedinger equation
\beq
(H-E)\Psi(E)=\Phi.
\eeq
In terms of the eigenfunctions $\Psi_n$ one has
\beq
\Psi(E)=
\sum_n\frac{\Psi_n\langle\Psi_n|\Phi
\rangle}{E_n-E}.
\eeq

The bootstrap conditions come from the requirement that 
in the gluon colour channel amplitudes
generated from their absorptive parts have the form corresponding to
the exchange of a single reggeized gluon. From this it
follows that
the spectrum of $H$ in this colour channel should contain
an eigenvalue $E_0=-\omega(q)$ here $\oq$ is the 
gluon Regge trajectory.
The corresponding eigenfunction
$\Psi_0$ is the gluon wave function:
\beq
(H+\omega(q))\Psi_0=0.
\eeq 
Eq. (7) is the first bootstrap condition.
The second bootstrap condition, introduced in [7], requires that
the residue of the PP amplitude 
at the gluon Regge pole correspond to its value
obtained from (1). In fact this is a condition which constrains
the form of the PPR vertex.

There is finally a third requirement. Namely, the amplitude generated
from its absorptive part should not contain admixture of other
eigenstates of the Hamiltonian different from the gluon. 
This requires that, as a function of reggeized gluons momenta, the
impact factor coincides with the gluonic wave function $\Psi_0$ up
to a coefficient which may depend on the momentum transfer $q$:
\beq
\Phi(q,q_1)=c(q)\Psi_0(q,q_1).
\eeq
This is the third bootstrap condition which we are going to discuss
in this note.

As pointed in [3], this third bootstrap condition is automatically 
satisfied in the NLO for the physical amplitude (1). Indeed, all
bootstrap conditions are satisfied in the leading order (LO).
As a result, the ratio
\[\langle\Psi_n|\Phi\rangle/\langle\Psi_0|\Phi\rangle\]
for states
$n\neq 0$ different from the one-gluon starts from the order $\alpha_s$.
From (3) we see that the relative contribution of states different from the
gluon involves a product of two such ratios for the projetile and target
and is therefore a correction of order $\alpha_s^2$, beyond the
NLO approximation.

However this is not true for the (unphysical) PR amplitudes,
which involve terms proportional to $\langle\Psi_n|\Phi\rangle$, of the
relative order $\alpha_s$. If (8) is not fulfilled in the NLO, then
in the NLO PR amplitudes will inevitably contain contributions
from states different from the one gluon state.
This does not spoil the unitarity for elastic amplitudes, but may
invalidate the unitarity for inelastic amplitudes, with production of
additional gluons, as discussed in [4].

The third bootstrap condition (8) implies that for any two physical 
particles the ratio of impact factors is independent of the reggeized
gluon momenta and so is a function of only momentum transfer $q$
\beq
\frac{\Phi_1(q,q_1)}{\Phi_2(q,q_1)}=F(q).
\eeq
This is a general statement, which does not depend on a particular
form of the gluon wave function.

A more particular form of this condition  can be written 
once the gluonic wave function is known. In [5] an ansatz for it was 
proposed (in the metric $\int d^2q_1'$ and for an Hermithean
Hamiltonian)
\beq
\Psi_0(q,q_1)=1/\sqrt{\ea\eb},\ \ q_1+q_2=q,
\eeq
where $\eq$ is a function which can be determined 
from the gluon trajectory $\oq$  by the equation
\beq
\oq=-\eq\langle\Psi_0|\Psi_0\rangle.
\eeq
(In fact the gluonic potential in the gluon colour channel is also
determined by $\eta$ to satisfy the first bootstrap condition (7), but
this  is irrelevant here). In the LO
\beq
\eta^{(0)}(q)=q^2/a,\ \ a=\frac{g^2N}{2(2\pi)^{D-1}}.
\eeq
In the NLO function $\eq$ was calculated in [3]. Later we shall
return to explicit form. 
One can try to verify if
(8) is true or not, with the gluonic wave function given by (10),

Presenting the impact factor as
\beq
\Phi(q,q_1)=\Phi^{(0)}(q,q_1)\big(1+\Delta(q,q_1)),
\eeq
where  term $\Delta$, of the order $\alpha_s$, is the NLO correction
one gets from (9) a requirement that the difference of $\Delta$'s
should be independent of $q_1$ for any two particles:
\beq
\Delta_1(q,q_1)-\Delta_2(q,q_1)=F_1(q).
\eeq
If we use the specific form (10) for the gluon wave function and
present
\beq
\eq=\eta^{(0)}(q)(1+\xi(q)),
\eeq
where again $\xi$ gives the NLO correction, then according to (8) the 
dependence of $\Delta(q,q_1)$
on $q_1$ has to be completely determined by that of $\eq$
\beq
\Delta(q,q_1)=-(1/2)(\xa+\xb)+\zeta(q),\ \ q_1+q_2=q,
\eeq
where $\zeta$ coming from $c$ in (9) is an unknown function of $q$.

Note that in Eqs. (14) and (16) it is assumed that the impact
factor is different from zero in the LO. In fact in the NLO appear
helicity changing terms, absent in the LO. For such terms the
third bootstrap condition reduces to the requirement that they
should be proportional to the LO gluon function, that is,
independent of the gluonic momenta.

It is these requirements that we are going to check for the impact factors
calculated in [1,2]. We restrict ourselves to a simpler case of
massless quarks.

\section{The gluon impact factor in the gluon channel}
In this and following sections we  reproduce the results of
[1,2] for the NLO impact factors in the massless quark case,
projected  onto the gluon colour 
channel in slightly different notation adapted to our purpose.
The gluon NLO impact factor  $\Phi^{(1)}_g$ 
consists of three terms
\beq
\Phi_g^{(1)}=\Phi_{g1}^{(1)}+\Phi_{g2}^{(1)}+\Phi_{g3}^{(1)}
\eeq
corresponding to the one-gluon,  $q\bar q$ and two-gluon
intermediate states respectively. Their explicit form is given
by the following expressions. The part from the one-gluon state
is
$$
\Phi_{g1}^{(1)}(q,q_1)=
\beta\Phi_{g}^{(0)}N
\biggl[ q_1^{2\epsilon}\ln\left( \frac{s_0}{q_1^2} \right)+
 q_2^{2\epsilon}\ln\left( \frac{s_0}{q_2^2} \right)
 + 
\biggl( \frac{2}{\epsilon} - \frac{(11 + 9\epsilon)}{2(1 +
2\epsilon)(3 + 2\epsilon)}
$$
$$
+ \frac{n_f}{N}\frac{(1 + \epsilon)(2 + \epsilon) - 1}{(1 + \epsilon)(1 +
2\epsilon)(3 + 2\epsilon)} + \psi(1) + \psi(1 - \epsilon) - 2\psi(1 + \epsilon)
\biggr)(q_1^{2\epsilon}+q_2^{2\epsilon})
\biggr]
$$
\begin{equation}\label{74}
+ \beta N(\chi (q_1)+\chi (q_2))\frac{2\epsilon}{(1 + \epsilon)(1 + 
2\epsilon)(3 + 2\epsilon)}\left( 1 + \epsilon -
\frac{n_f}{N} \right).
\end{equation}
Here $\Phi^{(0)}_g$ is the impact factor in the LO
\beq
\Phi^{(0)}_g=ig^2(e'e)_{\perp}\frac{\sqrt{N}}{2}T^a,
\eeq
with $e$ and $e'$ the gluon polarization vectors,  $T$ its colour vector
and $N$ and $n_f$ the numbers of colours and flavours.
Function $\chi$ is defined as
\beq
\chi(q_1)=g^2i\frac{\sqrt{N}}{2}T^aq_1^{2\epsilon}(e'q_1)(eq_1)_/q_1^2
\eeq 
and the coefficient is
\beq
\beta=\frac{g^2}{(4\pi)^{2+\epsilon}}\frac{\Ga(-\epsilon)\Ga^2(1+\epsilon)}
{\Ga(1+2\epsilon)},
\eeq
where $D=4+2\epsilon$ is the dimension used to calculate the 4-dimensional
divergent integrals.
The part coming from the $q\bar q$ state is
$$
\Phi_{g2}^{(1)}(q,q_1)=
-\beta\Phi_g^{(0)}n_f\frac{2(1 + \epsilon)^2 + \epsilon}
{(1 + \epsilon)(1 + 2\epsilon)(3 + 2\epsilon)}
(q_1^{2\epsilon}+q_2^{2\epsilon}-q^{2\epsilon})$$
\beq
+\beta n_f \frac{2\epsilon}{(1 + \epsilon)(1 + 2\epsilon)(3 + 2\epsilon)}
(\chi (q_1)+\chi (q_2)-\chi(q)).
\eeq
Finally the two-gluon intermediate state generates a contribution
$$
\Phi_ {g3}^{(1)}(q,q_1)=
-\beta N\Phi_g^{(0)}\Big[ 
 q_1^{2\epsilon}\ln\left( \frac{s_0}{q_1^2}\right)+
 q_2^{2\epsilon}\ln\left( \frac{s_0}{q_2^2}\right)-
 q^{2\epsilon}\ln\left( \frac{s_0}{q^2}\right)+
$$
$$
\biggl( \frac{3}{2\epsilon}
- \frac{(11 + 8\epsilon)}{(1 + 2\epsilon)(3 + 2\epsilon)} - \psi(1 + 2\epsilon)
- \psi(1 + \epsilon) + \psi(1 - \epsilon) + \psi(1) \biggr)
(q_1^{2\epsilon}+q_2^{2\epsilon}-q^{2\epsilon}) 
$$
\beq
+ \frac{1}{2}\biggl( \frac{1}{\epsilon} + 2\psi
(1 + 2\epsilon) - 2\psi(1 + \epsilon) + 2\psi(1 - \epsilon) - 2\psi(1) \biggr)
q^{2\epsilon} - \epsilon K_1\Big]  
-\beta N\frac{2\epsilon}{(1 + 2\epsilon)(3 + 2\epsilon)}
(\chi (q_1)+\chi (q_2)-\chi(q)),
\eeq
where, as an expansion in powers of $\epsilon$,
\beq
K_1 = \frac{1}{2}q^{2\epsilon}\biggl[
\frac{1}{\epsilon^2}\biggl( 2 - \left( \frac{q_1^2}{ q^2}
\right)^{\epsilon} - \left( \frac{ q_2^2}{ q^2} \right)
^{\epsilon} \biggr) + 4\psi^{\prime\prime}(1)\epsilon + \ln\left( \frac
{q_1^2}{q^2} \right)\ln\left( \frac{ q_2^2}{ q^2} \right) \biggr]~.
\eeq

Summing the three contribution we find the final NLO gluon impact factor as
\[
\Phi_{g}^{(1)}(q,q_1)=
\beta\Phi_g^{(0)}N\biggl[(q_1^{2\epsilon}+q_2^{2\epsilon})
\biggl(\frac{1}{2\epsilon}+\frac{11+7\epsilon}{2(1+2\epsilon)(3+2\epsilon)}-
\ps (1+\epsilon)+\ps (1+2\epsilon)\biggr)+\epsilon K_1(q,q_1)+\]\[
q^{2\epsilon}\biggl(\log\frac{s_0}{q^2}+\frac{1}{\epsilon}-
\frac{11+8\epsilon}{(1+2\epsilon)(3+2\epsilon)}-2\ps(1+2\epsilon)+2\ps(1)
\biggr)\biggr]-\]\[
-\beta\Phi_g^{(0)}n_f\Big[(q_1^{2\epsilon}+q_2^{2\epsilon})
\frac{1+\epsilon}{(1+2\epsilon)(3+2\epsilon)}-
q^{2\epsilon}\frac{2(1 + \epsilon)^2 + \epsilon}
{(1 + \epsilon)(1 + 2\epsilon)(3 + 2\epsilon)}\Big]\]\beq
+\beta\chi(q)\biggl[-n_f\frac{2\epsilon}{(1+\epsilon)
(1+2\epsilon)(3+2\epsilon)}
+ N\frac{2\epsilon}{(1+2\epsilon)(3+2\epsilon)}\biggr].
\eeq

\section{The quark impact factor in the gluon colour channel}
The quark NLO impact factor is a sum of two contributions coming from the
quark intermediate state and the quark-gluon intermediate state. 
The contribution from the quark intermediate state can be written as
\[
\Phi_{q1}^{(1)}/(\beta\Phi_q^{(0)})=
N q_1^{2\epsilon}\ln\left(\frac{s_0} {q_1 ^2}\right)+
N q_2^{2\epsilon}\ln\left(\frac{s_0} {q_2 ^2}\right)+ 
(q_1^{2\epsilon}+q_2^{2\epsilon}) 
\left\{
- n_f \frac{(1+\epsilon)}{(1+2\epsilon)(3+2\epsilon)}
-\frac{1}{N} \left[ \frac{1}{\epsilon(1+2\epsilon)} + \frac{1}{2} \right]
\right.
\]
\beq
\left.
+ N \left[\psi(1-\epsilon) - 2 \psi(\epsilon) + \psi(1)
+ \frac{1}{4(1+2\epsilon)(3+2\epsilon)} - \frac{1}{\epsilon(1+2\epsilon)}
- \frac{7}{4(1+2\epsilon)}\right]\right\}. 
\eeq
Here $\Phi_q^{(0)}$ is the LO quark impact factor in the gluon 
colour channel
\beq \Phi_q^{(0)}=-ig^2\frac{\sqrt{N}}{2}t\delta_{\lambda'\lambda},
\eeq
with $t$ the quark colour and $\lambda$'s its helicities.
The contribution from the quark-gluon intermediate state can be written as
\[
\Phi_{q2}^{(1)}(q,q_1)/(\beta\Phi_q^{(0)})=-\frac{N}{2}
\Big[ - 4 q^{2\epsilon} \biggl( \frac{1}{2}\ln\left(\frac{s_0}
{ q^2}\right) + \psi(1) - \psi(1+2\epsilon)
- \frac{3}{4(1+2\epsilon)} \biggr)\]
\[
 + (q_1^{2\epsilon}+q_2^{2\epsilon})
 \biggl(
-\frac{1}{\epsilon}  - \frac{3}{1+2\epsilon}
+ 2 \psi(1-\epsilon) - 2 \psi(1+2\epsilon) + 2\psi(1) - 2 \psi(\epsilon) 
\biggr)\] 
\beq
+ 2  q_1^{2\epsilon} \ln\left(\frac{s_0}{ q_1^2}\right)
+ 2  q_2^{2\epsilon} \ln\left(\frac{s_0}{ q_2^2}\right) 
- 2\epsilon K_1 \Big]
+\frac{1}{N}\biggl[\frac{1}{\epsilon (1+2\epsilon)}+\frac{1}{2}\biggr]
(q_1^{2\epsilon}+q_2^{2\epsilon}-q^{2\epsilon}). 
\eeq
Summing these two terms we find the quark NLO impact factor in the gluon 
channel as
\[
\Phi_q^{(1)}(q,q_1)=\beta\Phi_{q}^{(0)}
\biggl\{(q_1^{2\epsilon}+q_2^{2\epsilon})
\biggl[N\left(
\frac{1}{2\epsilon}+\frac{11+7\epsilon}{2(1+2\epsilon)(3+2\epsilon)}
-\ps(1+\epsilon)+
\ps(1+2\epsilon)\right)
\]\[-n_f\frac{1+\epsilon}{(1+2\epsilon)(3+2\epsilon)}
\biggr]+N\epsilon K_1(q,q_1)+\]\beq
q^{2\epsilon}\Big[N\left(\log\frac{s_0}{q^2}-\frac{3}{2(1+2\epsilon)}+2\ps(1)-
2\ps(1+2\epsilon)\right)-\frac{1}{N}\left(\frac{1}{\epsilon (1+2\epsilon)}
+\frac{1}{2}\right)\biggr\}.
\eeq

\section{The gluonic wave function and the NLO bootstrap}
Inspecting Eqs. (25) and (29) we conclude that requirement (14)
following from the third bootstrap condition is fulfilled.
Indeed the parts of the NLO correction to the ratio $\Phi/\Phi^{(0)}$
which depend on the gluonic momenta $q_1$ and $q_2$ are identical in
the gluon and quark impact factors. If a stronger condition (8)
is obeyed, this allows to find the gluon wave 
function in the NLO:
\beq
\Psi_0(q,q_1)=1+
a(q_1^{2\epsilon}+q_2^{2\epsilon})+\beta N\epsilon K_1(q,q_1),
\eeq
where
\beq a=\beta\biggl[N\left(
\frac{1}{2\epsilon}+\frac{11+7\epsilon}{2(1+2\epsilon)(3+2\epsilon)}
-\ps(1+\epsilon)+
\ps(1+2\epsilon)\right)
-n_f\frac{1+\epsilon}{(1+2\epsilon)(3+2\epsilon)}
\biggr]
\eeq
and $\beta$ and $K_1$ are given by (21) and (24) respectively.
Of course, arbitrary NLO terms  depending only on $q$ can be
added to (30), but this only influences the normalization of the
wave function. It is remarkable that the gluon wave function remains
infrared finite also in the NLO. Indeed in the limit 
$\epsilon\rightarrow 0$ one gets from (30) and (31) (dropping all terms
independent of $q_1$ and $q_2$)
\beq
\Psi_0=1-\frac{g^2}{32\pi^2}\Big[\left(\frac{11}{3}N-\frac{2}{3}n_f\right)
(\ln q_1^2 +\ln q_2^2)+
N(\ln q_1^2\ln q_2^2-
\ln q^2\ln q_1^2-\ln q^2\ln q_2^2)\Big].
\eeq

It is instructive to compare (30) with the form (16) which follows 
from the anzatz (10). The part of $\xi$ which comes from the quarks
(that is, proportional to $n_f$) was found in [3] to be
\beq
\xi(q)=2n_f\beta q^{2\epsilon}\frac{1+\epsilon}
{(1+2\epsilon)(3+2\epsilon)}.
\eeq
From (30) and (31)  we observe that the quark contribution to the
wave function is exactly given by (16). So the ansatz (10) works
in the quark sector. However, if (8) is true, it does not
work in the gluonic sector. The part of the gluon
function coming from gluon intermediate states contains function
$K_1(q,q_1)$, which cannot be represented in a simple form (16),
since it mixes dependence on $q,\ q_1$ and $q_2$. 
In the limit $\epsilon\rightarrow 0$ ansatz (10) gives only the first term 
in the
square brackets in (32) (linear in logarithms)
the second term, quadratic in the logarithms, missing.

Note that the term linear in logarithms, in fact, provides for the running 
in the coupling. Multiplying (32) by $g^2$ one presents the gluon
wave function up to the NLO in the form
\beq
\Psi_0(q,q_1)=g^2(q_1q_2)\left(1-\frac{g^2(q_1q_2)N}{32\pi^2}
(\ln q_1^2\ln q_2^2-
\ln q^2\ln q_1^2-\ln q^2\ln q_2^2)\right).
\eeq
where
\beq
g(q_1q_2)=g^2\Big[1-\frac{g^2}{16\pi^2}\left(\frac{11}{3}N-\frac{2}{3}
n_f\right)\ln q_1q_2\Big]
\eeq
is a running coupling constant at scale $q_1q_2$.
The running of the coupling and scale are correctly reproduced by ansatz (10).
However corrections of the 2nd order in the running coupling are missing.

Knowledge of the NLO wave function of the gluon allows to
write the main bootstrap condition (the ``first''one) in the
NLO as an identity to be satisfied by the NLO potential at
all $q$ and $q_1$, similar to its form in the LO:
\beq
(H^{(0)}+\omega^{(0)})\Psi_0^{(1)}+
(H^{(1)}+\omega^{(1)})\Psi_0^{(0)}=0.
\eeq
Here 
\beq
H=-\omega(q_1)-\omega(q_2)-V_{12}
\eeq
is the 2-gluon Hamiltonian in the gluon colour channel. Upper indeces
denote the LO (0) and NLO (1). In a more explicit form (36) reads:
\[
\int\frac{d^2q'_1}{\qpa\qpb}V_{12}^{(1)}(q_1,q_2|q'_1,q'_2)=
\omega^{(1)}(q)-\omega^{(1)}(q_1)-\omega^{(1)}(q_2)+\]\[
\Big(\omega^{(0)}(q)-\omega^{(0)}(q_1)-\omega^{(0)}(q_2)\Big)
\Big(a(q_1^{2\epsilon}+q_2^{2\epsilon})+\beta N\epsilon K_1(q,q_1)\Big)-
\]\beq
\frac{g^2N}{16\pi^3}\int\frac{d^2q'_1}{\qpa\qpb}
\left(\frac{q_1^2\qpb+q_2^2\qpa}{(q_1-q'_1)^2}-q^2\right)
\Big(a({q'_1}^{2\epsilon}+{q'_2}^{2\epsilon})+
\beta N\epsilon K_1(q,q'_1)\Big).
\eeq
Since the part of the NLO potential coming from the quark agrees
with the  anzatz (10), as found in [3], we expect that this
part satisfies (38). For the still unknown gluonic part Eq. (38)
presents a highly non-trivial condition to be satisfied at all $q$ and
$q_1$.

Note that considering (36) as an equation for the NLO part $\Psi_0^{(1)}$
one evidently finds a requirement that the inhomogeneous (2nd) term 
should be orthogonal to the solution of the homogeneous equation,
which is $\Psi_0^{(0)}=1$. So one gets a condition
\beq
\langle \Psi_0^{(0)}|H^{(1)}+\omega^{(1)}(q)|\Psi_0^{(0)}\rangle,
\eeq
which was first obtained in [7]. Relation (39) also gives  a condition
to be satisfied by the NLO potential. However it is much less
stringent as compared to (38), since it is only an identity in $q$,
the dependence on $q_1$ having been integrated out.
\section {Acknowledgments.}
The authors express their deep gratitude to Prof. G.Venturi for his
constant interest in this work. They  thank the INFN  and Physics
Department of Bologna University for their hospitality and
financial support. G.P.V. gratefully acknowledges the financial support
from the Alexander von Humboldt Stiftung.
\section{References.}
\noi 1.V.S.Fadin, R.Fiore, M.I.Kotsky and A.Papa, 
preprint BUDKERINP/99-61, UNICAL-TH 99/3 (hep-ph/9908264).\\
2.V.S.Fadin, R.Fiore, M.I.Kotsky and A.Papa, 
preprint BUDKERINP/99-62, UNICAL-TH 99/ (hep-ph/9908265).\\
3. M.A.Braun and G.P.Vacca, Phys. Lett. {\bf 454} (1999) 319.\\
4. L.N.Lipatov, Yad. Fiz. {\bf 23} (1976) 642; \ \
 J.Bartels, Nucl. Phys. {\bf B151} (1979) 293.\\
5. M.A.Braun, Phys. Lett. {\bf B345} (1995) 155; {\bf B348} (1995) 190.\\
6. V.S.Fadin, R.Fiore and A.Papa, Phys. Rev. {\bf D 60} (1999) 074025.\\
7. V.S.Fadin, R.Fiore, Phys. Lett. {\bf B 440} (1998) 359.

 \end{document}